\documentclass[preprint,12pt]{elsarticle}

\usepackage{geometry}
\usepackage{amssymb}
\usepackage{amsmath}
\usepackage{amsfonts}
\usepackage{amssymb}
\usepackage[version=4]{mhchem}
\usepackage{stmaryrd}

\usepackage[colorlinks,
            linkcolor=blue,
            anchorcolor=blue,
            citecolor=blue]{hyperref}

\usepackage{booktabs}
\usepackage{multirow}

\journal{}

\begin{document}

\begin{frontmatter}



\title{A Special Lorentz Violation Model and a Special Rainbow Function}


\author[inst1]{Jinwen Hu}
\ead{200731890025@whu.edu.cn}

\author[inst1]{Huan Hu}

\affiliation[inst1]{organization={Department of physics and technology},
            addressline={Wuhan university},
            city={Wuhan},
            postcode={430060},
            country={China}}

\begin{abstract}
In order to characterize the common feature of the general Lorentz violation models that the local speed of light is variable at ultrahigh energy scale, we introduced a parameter $n$ to characterize the variation of the speed of light between inertial systems. And in order not to violate some fundamental principles and experiments' results, some constraints were imposed on $n$. Then a coordinate transformation with the parameter $n$, which meets the symmetry of inertial systems, was naturally obtained.
But just to satisfy the symmetry of inertial systems, the expression for $n$ cannot be determined. Inspired by the idea of DSR model, we then discussed a specific expression for $n$ that makes the particle's energy have a limit rather than be infinite derived from the Lorentz model. We found that the modified dispersion relation corresponding to the specific expression for $n$ is deeply associated with the general formula of dispersion relation from the DSR. The motivation of introducing such a parameter $n$ was also discussed.
Finally, we discussed the possible LIV effects from astrophysical observations in GRB events and photon annihilation events. Using the data from GRB 190114C we investigated the special dispersion relation obtained in this paper and found that it appears to fit these data better than the two models corresponding to the first or second order approximation of the general formula of dispersion relation. And similar to the other Lorentz violation models corresponding to the subluminal case, the modified dispersion relation obtained in this paper can also result in increasing the energy threshold of photon annihilation reaction, which is expected to be verified by the future energetic photons events.
\end{abstract}




\end{frontmatter}


\section{Introduction}\label{sec1}
It is well known that the Special Relativity and General Relativity have already made great achievements, but at the same time the Lorentz violating models are also of some astrophysical interest. In the past few decades the scientific community has shown an intense interest in the theories that contained and investigated the breakdown of Lorentz symmetry in many scenarios \cite{bib01,bib02,bib03,bib04,bib05} and also the so-called Deformed Special Relativity (DSR) \cite{bib06}. For example, a common feature of semi-classical approaches to quantum gravity is the violation of Lorentz symmetry due to a deviation from the usual relativistic dispersion relation caused by a redefinition of the physical momentum and physical energy at the Planck scale. And one of the intriguing predictions among various quantum gravity theories, such as loop quantum gravity, non-commutative geometry, and string theories, is the existence of a maximum measurable energy that nears the Planck energy.

A possible deformed Lorentz symmetry driven by a transformation in momentum space was initially presented by J. Magueijo and L. Smolin \cite{bib07,bib08}. In their work, the Planck energy scale was assumed to be a second constant between inertial systems besides the speed of light, and correspondingly, a general formula of dispersion relation could be constructed as ( $\operatorname{set} \boldsymbol{c}=1$ )
\begin{equation}
    \label{eq1}
    \left[1+\chi_{1}\left(\frac{E}{E_{L V}}\right)^{1}+\chi_{2}\left(\frac{E}{E_{L V}}\right)^{2}+O\left(\frac{E}{E_{L V}}\right)^{3}\right] E^{2}-\boldsymbol{p}^{2}=m_{0}^{2}
\end{equation}
where $E$ denotes the total energy of the particle, $m_{0}$ denotes the rest mass, $\boldsymbol{p}$ is the momentum, and $E_{\mathrm{LV}}$ denotes the energy scale at which Lorentz violating effects become strong, the couplings $\chi_{s}(s=1,2)$ are determined by the experiments.

Eq. (\ref{eq1}) is also called the ``rainbow model", which states that the space-time background depends on the energy of a tested particle. Due to the large scale of $E_{\mathrm{LV}}$, the variation of $\boldsymbol{c}$ is extremely small at low energy scale that it is very difficult to measure by ordinary experiments. But a feasible approach to solve this problem is to detect photons from the astrophysical objects, such as the Gamma Ray Burst (GRB) events. In recent years, many physicists have used the rainbow model to study the variation of the speed of light in GRB events \cite{bib09,bib10,bib11,bib12,bib13,bib14,bib15,bib16}. For example, with the first order approximation of Eq.(\ref{eq1}), Xu \cite{bib16,bib17} analyzed the GRB 160509A event and they claimed that $E_{\mathrm{LV}} \approx 3.6 \times 10^{17} \mathrm{GeV}$.

In this paper, we introduced a new parameter into the coordinate transformation between inertial systems to characterize the violation of Lorentz model, and by defining a special expression for the introduced parameter we obtained a special case of Eq. (\ref{eq1}) that the inertial systems remain to be equivalent but the particle's total energy could have a limit. And thus the paper was organized as follows. In Sect. \ref{sec2}, we raised a parameter $n$ to characterize the variation of the speed of light between inertial systems, and with the constraints from some fundamental principles, we obtained a general coordinate transformation, which meets the symmetry of inertial systems. In Sect. \ref{sec3}, inspired by the idea of rainbow model, we construct a specific expression for $n$ to make the particle's energy have a limit rather than be infinite derived from the Lorentz model. In Sect. \ref{sec4}, we investigated the dispersion relation derived in this paper at ultrahigh energy scale, and obtained a special rainbow function. In Sect. \ref{sec5} we stated the motivation of introducing such a parameter $n$. In Sect. \ref{sec6} we discussed the LIV
effects from astronomical observations in GRB events and photon annihilation events, to investigate the special dispersion relation obtained in this paper. In Sect. \ref{sec7} we summarized the paper.

\section{Variable speed of light}\label{sec2}
Speed of light is considered to be a constant in all inertia systems in special relativity, which leads to the Lorentz invariance. However, as noted above, in order to unify theories in different fields, such as the string theory, which attempts to unify quantum theory and general relativity, or to solve some problems in the theory itself, such as the extension of Standard Model, a possible violation of Lorentz invariance is proposed, for example, the deformed or doubly special relativity (DSR) is just this class of theories, which predict that at ultrahigh energy scale, the speed of light is obviously variable.

In many literature, such as Ref. \cite{bib15,bib16}, the authors used the rainbow model to study the Lorentz invariance violation (LIV) effects, and present that the speed of light maybe associated with the photon's energy. However, here in order to discuss the general case of the speed of light related to the violation of Lorentz invariance, we raised a general hypothesis that: For a light source in vacuum, when it moves at a velocity $\boldsymbol{v}$ relative to an observer in vacuum, then the observed (by the observer in vacuum) speed of light emitted by the light source is $n \boldsymbol{c}$, where $n$ is a dimensionless quantity, $\boldsymbol{c}$ is the speed of light in vacuum. Obviously, in order not to violate some fundamental principles and experiments' results, we should impose some constraints on the parameter $n$ as follows

\begin{enumerate}[(1)]
  \item Firstly, as stated in Einstein's special relativity, using the speed of light it should be possible to define the (proper) time (note that the only time that makes sense in special relativity is the proper time measured by the clocks carried by physical observers) in the whole space with a prescribed clock synchronization. That is, for a specific inertial system, using the speed of light emitted by a light source fixed in the specific inertial system, one can calibrate the clock fixed in the inertial system to synchronize. So it requires that
\end{enumerate}
\begin{equation}
    \label{eq2}
    n(\boldsymbol{v}=0, \boldsymbol{c})=1
\end{equation}
\begin{enumerate}[(2)]
  \setcounter{enumi}{1}
  \item Secondly, in order to satisfy the general concept that space-time is uniform and space is isotropic, it requires that $n(\boldsymbol{v}, \boldsymbol{c})$ is independent of the direction of vector $\boldsymbol{v}$ and $\boldsymbol{c}$, namely,
\end{enumerate}
\begin{equation}
    \label{eq3}
    n(v, c)=n(-v, c)=n(v,-c)=n(-v,-c)
\end{equation}
\begin{enumerate}[(3)]
  \setcounter{enumi}{2}
  \item In addition, here we also follow the principle that all the inertial systems are equivalent.
\end{enumerate}

Based on the above assumption on the speed of light, next we will derive the coordinate transformation between the two inertial systems $S(\boldsymbol{x}, \boldsymbol{y}, \boldsymbol{z}, t)$ and $S^{\prime}\left(\boldsymbol{x}^{\prime}, \boldsymbol{y}^{\prime}, \boldsymbol{z}^{\prime}, t^{\prime}\right)$. And here we assume that $S^{\prime}$ is moving at a velocity $\boldsymbol{v}$ relative to $S$.

Firstly, for simplicity, we assume that the three spatial coordinates of the two coordinate systems are parallel to each other, and the direction of $\boldsymbol{v}$ is along the $\boldsymbol{x}$-axis or $\boldsymbol{x}^{\prime}$-axis. Then it hints that $\boldsymbol{y}=\boldsymbol{y}^{\prime}, \boldsymbol{z}=\boldsymbol{z}^{\prime}$.

Secondly, since space-time is uniform, the coordinate transformation between $S$ and $S^{\prime}$ should be in a linear form, and we assume that
\begin{equation}
    \label{eq4}\boldsymbol{x}=\gamma\left(\boldsymbol{x}^{\prime}+\boldsymbol{v} t^{\prime}\right)
\end{equation}
Where $\gamma=\gamma(\boldsymbol{v}, \boldsymbol{c})$ is a proportionality factor.

The form of Eq. (\ref{eq4}) should be invariant under the time reversal symmetry operation. Thus we have
\begin{equation}
    \label{eq5}
    \gamma(\boldsymbol{v}, \boldsymbol{c})=\gamma(-\boldsymbol{v},-\boldsymbol{c})
\end{equation}

Note that in above we didn't distinguish the direction of vector $\boldsymbol{v}$ and $\boldsymbol{c}$, i.e., the direction of vector $\boldsymbol{v}$ and $\boldsymbol{c}$ maybe along the positive $\boldsymbol{x}$-axis (or $\boldsymbol{x}^{\prime}$-axis) or along the negative $\boldsymbol{x}$-axis (or $\boldsymbol{x}^{\prime}$-axis). If we distinguish the direction of vector $\boldsymbol{v}$ and $\boldsymbol{c}$ by the positive and negative signs, we will obtain four different combinations, i.e., $(v, c),(v$, $-c),(-v, c),(-v,-c)$. Based on Eq. (\ref{eq5}) we can obtain $\gamma(v, c)=\gamma(-v,-c), \gamma(-v, c)=\gamma(v,-c)$.

On the other hand, if the light signal is emitted by the light source at the moment that the origin of $S$ and $S^{\prime}$ are coincides, then based on the above assumption on the speed of light, we will obtain
\begin{equation}
    \label{eq6}
\left\{\begin{array} { l }
{ \boldsymbol{x} ^ { 2 } + \boldsymbol{y} ^ { 2 } + \boldsymbol{z} ^ { 2 } = ( \boldsymbol{c} t ) ^ { 2 } } \\
{ \boldsymbol{x} ^ { \prime 2 } + \boldsymbol{y} ^ { \prime 2 } + \boldsymbol{z} ^ { \prime 2 } = ( n \boldsymbol{c} t ^ { \prime } ) ^ { 2 } } \\
{ \boldsymbol{y} = \boldsymbol{y} ^ { \prime } = 0 } \\
{ \boldsymbol{z} = \boldsymbol{z} ^ { \prime } = 0 }
\end{array} \quad \text { or } \quad \left\{\begin{array}{l}
\boldsymbol{x}^{\prime 2}+\boldsymbol{y}^{\prime 2}+\boldsymbol{z}^{\prime 2}=\left(\boldsymbol{c} t^{\prime}\right)^{2} \\
\boldsymbol{x}^{2}+\boldsymbol{y}^{2}+\boldsymbol{z}^{2}=(n \boldsymbol{c} t)^{2} \\
\boldsymbol{y}=\boldsymbol{y}^{\prime}=0 \\
\boldsymbol{z}=\boldsymbol{z}^{\prime}=0
\end{array}\right.\right.
\end{equation}

The first formula of Eq. (\ref{eq6}) represents that when the light source is fixed in $S$, then for the observer in $S$, the observed speed of light is $\boldsymbol{c}$ (based on Eq. (\ref{eq2})), while for another observer in $S^{\prime}$, the observed speed of light is $n\boldsymbol{c}$.

Similarly, since $S$ and $S$ are equivalent, when the light source is fixed in $S$, then for the observer in $S$, the observed speed of light is $\boldsymbol{c}$, while for another observer in $S$, the observed speed of light is $n\boldsymbol{c}$, which case corresponds to the second formula of Eq. (\ref{eq6}).

Now we will solve the expression for $\gamma$. From Eq. (\ref{eq4}) and Eq. (\ref{eq6}), we can obtain that
\begin{equation}
    \label{eq7}
\left\{\begin{array} { l }
{ \boldsymbol { x } = \gamma ( \boldsymbol { x } ^ { \prime } + \boldsymbol{v} t ^ { \prime } ) } \\
{ \boldsymbol { x } ^ { \prime } = \gamma ( \boldsymbol { x } - \boldsymbol { v } t ) } \\
{ \boldsymbol { x } ^ { \prime } = n \boldsymbol{c} t ^ { \prime } } \\
{ \boldsymbol { x } = \boldsymbol { c } t }
\end{array} \quad \text { or } \quad \left\{\begin{array}{l}
\boldsymbol{x}^{\prime}=\gamma(\boldsymbol{x}-\boldsymbol{v} t) \\
\boldsymbol{x}=\gamma\left(\boldsymbol{x}^{\prime}+\boldsymbol{v} t^{\prime}\right) \\
\boldsymbol{x}^{\prime}=-\boldsymbol{c} t^{\prime} \\
\boldsymbol{x}=-n \boldsymbol{c} t
\end{array}\right.\right.
\end{equation}

Where the first formula corresponds to the first formula of Eq. (\ref{eq6}), and the second formula corresponds to the second formula of Eq. (\ref{eq6}). Note that the last two items in each formula of Eq. (\ref{eq7}) correspond to one of the two solutions of Eq. (\ref{eq6}). And here Eq. (\ref{eq5}) has been used.

Substituting the last two items into the first two items in each formula of Eq. (\ref{eq7}), it derives that
\begin{equation}
    \label{eq8}
\left\{\begin{array} { l }
{ \boldsymbol { c } t = \gamma ( n\boldsymbol { c } t ^ { \prime } +\boldsymbol {v} t ^ { \prime } ) = \gamma ( n \boldsymbol { c } +\boldsymbol {v} ) t ^ { \prime } } \\
{ n \boldsymbol { c } t ^ { \prime } = \gamma ( \boldsymbol { c } t - \boldsymbol {v} t ) = \gamma ( \boldsymbol { c } - \boldsymbol { v } ) t }
\end{array}  \text { or }  \left\{\begin{array}{l}
-\boldsymbol { c } t^{\prime}=\gamma(-n \boldsymbol { c } t-\boldsymbol {v} t)=-\gamma(n \boldsymbol { c }+\boldsymbol {v}) t \\
-n \boldsymbol { c } t=\gamma\left(-\boldsymbol { c } t^{\prime}+\boldsymbol {v} t^{\prime}\right)=-\gamma(\boldsymbol{c}-\boldsymbol{v}) t^{\prime}
\end{array}\right.\right.
\end{equation}

Multiplying the two left sides and the two right sides of Eq. (\ref{eq8}), we can obtain
the same equation
\begin{equation}
    \label{eq9}
    n\boldsymbol{c}^{2}=\gamma^{2}(n \boldsymbol{c}+\boldsymbol{v})(\boldsymbol{c}-\boldsymbol{v})
\end{equation}

Now the expression for $\gamma$ can be obtained based on Eq. (\ref{eq9}). And then it is easy to obtain the coordinate transformation between $S$ and $S^{\prime}$
\begin{equation}
    \label{eq10}
\left\{\begin{array}{l}
\boldsymbol{x}^{\prime}=\gamma(\boldsymbol{x}-\boldsymbol{v} t) \\
t^{\prime}=\gamma\left(t-\frac{\boldsymbol{v}}{k^{2}(\boldsymbol{v}, \boldsymbol{c})} \boldsymbol{x}\right)
\end{array}\right.
\end{equation}
where $\gamma(\boldsymbol{v}, \boldsymbol{c})=1 / \sqrt{1-\boldsymbol{v}^{2} / k^{2}}, k(\boldsymbol{v}, \boldsymbol{c})=\sqrt{n \boldsymbol{v} \boldsymbol{c}^{2} /(n \boldsymbol{c}-\boldsymbol{c}+\boldsymbol{v})}$.

From Eq. (\ref{eq10}) one can obtain that $k(v, c)=k(-v,-c), k(-v, c)=k(v,-c)$, and correspondingly, $\gamma(v, c)=\gamma(-v,-c), \gamma(-v, c)=\gamma(v,-c)$, which is consistent with Eq. (\ref{eq5}).

Keeping in mind that in Eq. (\ref{eq7}) we just used one of the two solutions of Eq. (\ref{eq6}) and then naturally obtained Eq. (\ref{eq9}), so what is the expression for $\gamma$ corresponding to the other solution of Eq. (\ref{eq6})? Here we approach the answer from another perspective. And at the same time in order to show the self-consistency of the above model, and as well as to better understand the transformation of Eq. (\ref{eq10}), we will reiterate the meaning of the speed of light in the next.

Based on Eq. (\ref{eq10}) one can obtain
\begin{equation}
    \label{eq11}
    \frac{d \boldsymbol{x}^{\prime}}{d t^{\prime}}=\frac{d \boldsymbol{x}-\boldsymbol{v}d t}{d t-\frac{\boldsymbol{v}}{k^{2}} d \boldsymbol{x}}=\frac{d \boldsymbol{x} / d t-\boldsymbol{v}}{1-\frac{\boldsymbol{v}}{k^{2}(\boldsymbol{v}, \boldsymbol{c})} \frac{d \boldsymbol{x}}{d t}}=f(\boldsymbol{v}, \boldsymbol{c})
\end{equation}

As stated above that the direction of vector $\boldsymbol{v}$ and $\boldsymbol{c}$ maybe along the positive $\boldsymbol{x}$-axis (or $\boldsymbol{x}^{\prime}$-axis) or along the negative $\boldsymbol{x}$-axis (or $\boldsymbol{x}^{\prime}$-axis), and if we distinguish the direction of vector $\boldsymbol{v}$ and $\boldsymbol{c}$ by the positive and negative signs, we will obtain four different combinations, i.e., $(v, c),(v,-c),(-v, c),(-v,-c)$. Then based on Eq. (\ref{eq11}), we have

Case 1: Note that the above we assumed that $S^{\prime}$ is moving at a velocity $\boldsymbol{v}$ relative to $S$. When the light source is fixed in $S$, then for the observer in $S$, the observed speed of light is $\boldsymbol{c}(=d \boldsymbol{x} / d t)$, while for the observer in $S^{\prime}$, it has
\begin{equation}
    \label{eq12}
    \left\{\begin{array}{l}
\text { when } \frac{d x}{d t}=c, \frac{d x^{\prime}}{d t^{\prime}}=f(v, c)=\frac{c-v}{1-\frac{v}{k^{2}(v, c)} c}=f(-v, c)=\frac{c+v}{1-\frac{-v}{k^{2}(-v, c)} c}=n c \\
\text { when } \frac{d x}{d t}=-c, \frac{d x^{\prime}}{d t^{\prime}}=f(v,-c)=\frac{-c-v}{1-\frac{v}{k^{2}(v,-c)}(-c)}=f(-v,-c)=\frac{-c+v}{1-\frac{-v}{k^{2}(-v,-c)}(-c)}=-n c
\end{array}\right.
\end{equation}

Case 2: Similarly, since $S$ and $S$ are equivalent, when the light source is fixed in $S^{\prime}$, then for the observer in $S$, the observed speed of light is $\boldsymbol{c}\left(=d \boldsymbol{x}^{\prime} / d t^{\prime}\right)$, while for the observer in $S$, it has (note that in this case the velocity of $S$ relative to the light source
or $S^{\prime}$ is $-\boldsymbol{v}$ )

\begin{equation}
    \label{eq13}
\left\{\begin{array}{l}
\frac{d x}{d t}=\frac{d \boldsymbol{x}^{\prime} / d t^{\prime}-(-\boldsymbol{v})}{1-\frac{-\boldsymbol{v}}{k^{2}(-\boldsymbol{v}, \boldsymbol{c})} \frac{d \boldsymbol{x}^{\prime}}{d t^{\prime}}}=f^{\prime}(\boldsymbol{v}, \boldsymbol{c}) \\
\text { when } \frac{d x^{\prime}}{d t^{\prime}}=c, \frac{d x}{d t}=f^{\prime}(v, c)=\frac{c+v}{1+\frac{v}{k^{2}(-v, c)} c}=f^{\prime}(-v, c)=\frac{c-v}{1+\frac{-v}{k^{2}(v, c)} c}=n c \\
\text { when } \frac{d x}{d t}=-c, \frac{d x}{d t}=f^{\prime}(v,-c)=\frac{-c+v}{1+\frac{v}{k^{2}(-v,-c)}(-c)}=f^{\prime}(-v,-c)=\frac{-c-v}{1+\frac{-v}{k^{2}(v,-c)}(-c)}=-n c
\end{array}\right.
\end{equation}

Eq. (\ref{eq12}) and Eq. (\ref{eq13}) can be expressed in a vector form
\begin{equation}
    \label{eq14}
\left\{\begin{array}{l}
\text { when } \frac{d \boldsymbol{x}}{d t}=\boldsymbol{c}, \frac{d \boldsymbol{x}^{\prime}}{d t^{\prime}}=n \boldsymbol{c} \\
\text { when } \frac{d \boldsymbol{x}^{\prime}}{d t^{\prime}}=\boldsymbol{c}, \frac{d \boldsymbol{x}}{d t}=n \boldsymbol{c}
\end{array}\right.
\end{equation}

Eq. (\ref{eq14}) shows all the meaning of the speed of light in inertial systems, and it also indicates that Eq. (\ref{eq10}) is just the solution of Eq. (\ref{eq6}).

Obviously, it can be seen that the forms of Eq. (\ref{eq10}) are similar to the Lorentz transformation, that is, replacing $\boldsymbol{c}$ in the Lorentz transformation with $k$ we can obtain Eq. (\ref{eq10}). Further, based on the similarity of the two forms, it is easy to prove that the Maxwell's equations are also covariant based on Eq. (\ref{eq10}).

Similar to the special relativity, based on Eq. (\ref{eq10}) we can obtain the line element of space-time
\begin{equation}
    \label{eq15}
    d s^{2}=-k^{2} d t^{2}+d \boldsymbol{x}^{2}
\end{equation}

Correspondingly, the particle's energy-momentum dispersion relation is
\begin{equation}
    \label{eq16}
E^{2}=\boldsymbol{p}^{2} k^{2}+E_{0}^{2}
\end{equation}
Where $E_{0}=m_{0} k^{2}$ denotes the particle's rest energy, $E=\gamma m_{0} k^{2}$ denotes the total energy, $\boldsymbol{p}=\gamma m_{0} \boldsymbol{v}$ denotes the particle's momentum.

\section{Particle's ``maximum energy"}\label{sec3}
As we know, in Lorentz model the particle's energy tends to be infinite when the particle's velocity is close to the speed of light, however, the idea of DSR or the rainbow model introduces a new constant (i.e., the Planck energy scale) as the energy limit of the particles. In above, as Eq. (\ref{eq10}) shown that if $n \equiv 1$ then Eq. (\ref{eq10}) returns to the Lorentz model. But here we would like to discuss another interesting case where $n$ is not always equal to 1 , as the general Lorentz violating models suggested \cite{bib01,bib02,bib03,bib04,bib05,bib06}.

Inspired by the idea of rainbow model, we found that Eq. (\ref{eq10}) implied that it is possible the particle's energy could have a limit. That is, based on Eq. (\ref{eq10}), the
space-time scaling factor is
\begin{equation}
    \label{eq17}
\gamma=\frac{1}{\sqrt{1-\boldsymbol{v}^{2} / k^{2}}}=\frac{1}{\sqrt{\frac{1-\boldsymbol{v} / \boldsymbol{c}}{n}\left(n+\frac{\boldsymbol{v}}{\boldsymbol{c}}\right)}}
\end{equation}

Eq. (\ref{eq17}) inspires us that, for the case of $\boldsymbol{v}=\boldsymbol{c}$, if we assume $n=0$ in this case, then it is possible that $\gamma$ or $E$ does not tend to be infinite, that is
\begin{equation}
    \label{eq18}
\lim _{v \rightarrow c} \frac{1-\boldsymbol{v} / \boldsymbol{c}}{n}=\text { const. }
\end{equation}

Remembering that $n$ has been constrained in Eq. (\ref{eq2}) and Eq. (\ref{eq3})
\begin{equation}
    \label{eq19}
\left\{\begin{array}{l}
n(\boldsymbol{v}=0, \boldsymbol{c})=1 \\
n(v, c)=n(v,-c)=n(-v, c)=n(-v,-c)
\end{array}\right.
\end{equation}

From a mathematical point of view, the set of function of $n$ satisfying Eq. (\ref{eq18}) and Eq. (\ref{eq19}) is large. For example, the next two functions of $n$ can be easily obtained
\begin{equation}
    \label{eq20}
\left\{\begin{array}{l}
n_{1}=\frac{1}{1-Q}\left(1-Q^{1-\boldsymbol{v}^{2} / \boldsymbol{c}^{2}}\right) \\
n_{2}=\sum_{j=1}^{j} a_{j}\left(1-\boldsymbol{v}^{2} / \boldsymbol{c}^{2}\right)^{j}
\end{array}\right.
\end{equation}
Where $Q$ and $a_{j}$ are constants, $j \in \mathrm{Z}+$ and $\sum_{j=1}^{j} a_{j}=1$. And here the form of $n_{1}$ is inspired by the l'Hôpital's rule, that is, for $\boldsymbol{v} \sim \mathbf{c}$,
\begin{equation}
    \label{eq21}
n_{1} \sim \frac{\ln Q}{Q-1}\left(1-\frac{\boldsymbol{v}^{2}}{\boldsymbol{c}^{2}}\right)
\end{equation}

However, at present it seems to be no physical evidence or rules for us to determine the function of $n$, but as we know there are many experiments restricting the violation of Lorentz model \cite{bib18,bib19,bib20,bib21,bib22,bib23,bib24,bib25,bib26,bib27}, so the expression for $n$ should be able to satisfy the previous experiments' results. For example, based on T. Alvager's experiment \cite{bib27}, if we choose $n_{1}$ as the function of $n$, then we can obtain that $Q<(1 / e)^{10^{6}}$, which indicates $Q \sim 0$. Correspondingly, as an example to show, figure 1 presents the curve of $n \sim \boldsymbol{v}$ when taking $Q=(1 / e)^{10^{6}}$.

But for $n_{2}$, for the same T. Alvager's experiment \cite{bib27}, we found it satisfy the experiment's result hardly unless we take $j>6$, which means there are at least five parameters involved this model to be determined.

Obviously, the form of $n_{1}$ is more simpler than $n_{2}$ for that there is just one parameter involved in $n_{1}$ to be determined, and more importantly, the property of $n_{1}$ (as can be seen from Fig. 1 that over large range of $v, n$ can remain to be close to 1 and at the same time when $\boldsymbol{v} \in[0, \boldsymbol{c}], n \geq 0$, which is necessary for the real world) can allow it not to violate the results of existing experiments at low or medium energy scale well.

In addition to the above two possible forms of $n$, one may also consider the other expressions for $n$ (such as the combination of $n_{1}$ and $n_{2}$ ), which may be more complicated than Eq. (\ref{eq20}), but here and in this paper, for the reasons mentioned above and as well as a matter of preference, we prefer to choose $n_{1}$ as the function of $n$ and discuss the effect caused by it in the later chapter. And the effects corresponding to the other forms of $n$ will be discussed in another paper.

\begin{figure}
    \centering
    \includegraphics[width=0.8\linewidth]{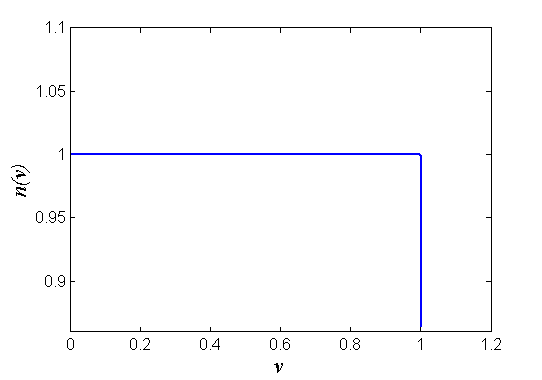}
    \caption{The curve of $n(\boldsymbol{v}) \sim \boldsymbol{v}$ when taking $Q=(1 / e)^{10^{6}} \quad($ set $\boldsymbol{c}=1)$. Since here it shows the global picture of the curve, a transition arc at the corner is too small to show.}
    \label{fig01}
\end{figure}

Thus, based on the first formula of Eq. (\ref{eq20}), both the space-time scaling factor and the particle's total energy have a limit, that is
\begin{equation}
    \label{eq22}
\begin{aligned}
& \lim _{v \rightarrow \boldsymbol{c}} \gamma=\lim _{\boldsymbol{v} \rightarrow \boldsymbol{c}} \frac{1}{\sqrt{1-\boldsymbol{v}^{2} / k^{2}}}=\lim _{\boldsymbol{v} \rightarrow \boldsymbol{c}} \frac{1}{\sqrt{\left(\frac{1-\boldsymbol{v} / \boldsymbol{c}}{n}\right)\left(n+\frac{\boldsymbol{v}}{\boldsymbol{c}}\right)}}=\sqrt{\frac{2 \ln Q}{Q-1}} \\
& \lim _{\boldsymbol{v} \rightarrow \boldsymbol{c}} E=\lim _{\boldsymbol{v} \rightarrow \boldsymbol{c}} \gamma m_{0} k^{2}=E_{Q G}=\frac{m_{0} \boldsymbol{c}^{2}}{[1-0.5(Q-1) / \ln Q]} \sqrt{\frac{2 \ln Q}{Q-1}}
\end{aligned}
\end{equation}

\section{A special rainbow function}\label{sec4}
It can be seen from Fig.1 that the modified particle's energy-momentum dispersion relation will return to the case in Lorentz model at low or medium energy. So next we will discuss the behavior of particles at ultrahigh energy scale.

When $\boldsymbol{v} \sim \boldsymbol{c}$, it can be obtained from Eq. (\ref{eq20}) that (set $\boldsymbol{c}=1$ )
\begin{equation}
    \label{eq23}
n=\frac{1}{1-Q}\left(1-Q^{1-\boldsymbol{v}^{2}}\right)=\frac{1}{1-Q}\left[1-Q^{(1+\boldsymbol{v})(1-\boldsymbol{v})}\right] \approx \frac{1}{1-Q}\left[1-Q^{2(1-\boldsymbol{v})}\right] \approx \frac{2 \ln Q}{Q-1}(1-\boldsymbol{v})
\end{equation}

Then
\begin{equation}
    \label{eq24}
\begin{aligned}
\frac{E}{E_{Q G}}& =\frac{m_0 k^2}{\sqrt{1-v^2 / k^2}}\left[\frac{m_0 c^2}{[1-0.5(Q-1) / \ln Q]} \sqrt{\frac{2 \ln Q}{Q-1}}\right]\\
& =\frac{[1-0.5(Q-1) / \ln Q]}{\sqrt{2 \ln Q /(Q-1)}} \frac{n \boldsymbol{v} /(n-1+\boldsymbol{v})}{\sqrt{(1-\boldsymbol{v})(1+\boldsymbol{v} / n)}} \\
& \approx \frac{[1-0.5(Q-1) / \ln Q]}{\sqrt{2 \ln Q /(Q-1)}} \frac{2 \ln Q /(Q-1) \boldsymbol{v}}{2 \ln Q /(Q-1)-1} \frac{1}{\sqrt{1-\boldsymbol{v}+\boldsymbol{v}(Q-1) /(2 \ln Q)}} \\
& \approx \frac{[1-0.5(Q-1) / \ln Q]}{\sqrt{2 \ln Q /(Q-1)}} \frac{2 \ln Q /(Q-1)}{2 \ln Q /(Q-1)-1} \frac{1}{\sqrt{(Q-1) /(2 \ln Q)+(1-\boldsymbol{v})}} \\
& \approx \frac{1}{\sqrt{2 \ln Q /(Q-1)}}\left[\sqrt{\frac{2 \ln Q}{Q-1}}-\frac{1}{2}\left(\frac{2 \ln Q}{Q-1}\right)^{3 / 2}(1-\boldsymbol{v})\right] \\
& =1-\frac{\ln Q}{Q-1}(1-\boldsymbol{v})
\end{aligned}
\end{equation}

Based on Eq. (\ref{eq24}) we can obtain
\begin{equation}
    \label{eq25}
\frac{\boldsymbol{v}}{\boldsymbol{c}}=1-\frac{Q-1}{\ln Q}+\frac{Q-1}{\ln Q} \frac{E}{E_{Q G}}
\end{equation}

Eq. (\ref{eq25}) shows that for an ultra-relativistic particle governed by Eqs. (\ref{eq26}) and (\ref{eq20}), its velocity is proportional to its energy.

Multiplying $m c^{2}$ on both sides of Eq. (\ref{eq25}), we can obtain
\begin{equation}
    \label{eq26}
\boldsymbol{p c}=m \boldsymbol{c}^{2}\left(1-\frac{Q-1}{\ln Q}+\frac{Q-1}{\ln Q} \frac{E}{E_{Q G}}\right)
\end{equation}
where $\boldsymbol{p}$ is the particle's momentum.

Note that since $Q \sim 0$, it is easy to prove that $E=m k^{2} \approx m c^{2}$ due to $k \approx c$. Then Eq. (\ref{eq26}) can be written as
\begin{equation}
    \label{eq27}
\boldsymbol{p c} \approx E\left(1-\frac{Q-1}{\ln Q}+\frac{Q-1}{\ln Q} \frac{E}{E_{Q G}}\right)
\end{equation}

On the other hand, in the framework of the DSR or rainbow model, for an ultra-relativistic particle, i.e., $\boldsymbol{v} \sim \boldsymbol{c}$, Eq. (\ref{eq1}) can be rewritten as
\begin{equation}
    \label{eq28}
p c \approx E \sqrt{1+\chi_{1}\left(\frac{E}{E_{L V}}\right)^{1}+\chi_{2}\left(\frac{E}{E_{L V}}\right)^{2}+O\left(\frac{E}{E_{L V}}\right)^{3}}
\end{equation}

Comparing Eq. (\ref{eq27}) with Eq. (\ref{eq28}), we can obtain
\begin{equation}
    \label{eq29}
\left\{\begin{array}{l}
\chi_{1} \frac{1}{E_{L V}}=2\left(1-\frac{Q-1}{\ln Q}\right)\left(\frac{Q-1}{\ln Q}\right) \frac{1}{E_{Q G}} \\
\chi_{2} \frac{1}{E_{L V}{ }^{2}}=\left(\frac{Q-1}{\ln Q}\right)^{2} \frac{1}{E_{Q G}{ }^{2}}
\end{array}\right.
\end{equation}

Eq. (\ref{eq29}) shows that the dispersion relation derived in this paper is deeply associated with the general formula of dispersion relation raised in DSR. But the two
``maximum energy", i.e., $E_{LV}$ and $E_{QG}$, are fundamentally different. That is, $E_{LV}$ raised in DSR is considered to be a constant between inertial systems, while $E_{QG}$ is not a constant but depends on the particle's rest mass.

Further, based on Eq. (\ref{eq29}) we can obtain that
\begin{equation}
    \label{eq30}
\frac{\chi_{1}^{2}}{\chi_{2}}=4\left(1-\frac{Q-1}{\ln Q}\right)^{2}
\end{equation}

Since $Q \sim 0$, Eq. (\ref{eq30}) can be simplified as
\begin{equation}
    \label{eq31}
\frac{\chi_{1}^{2}}{\chi_{2}} \approx 4
\end{equation}
which presents a special case of Eq. (\ref{eq1}).

\section{The motivation of introducing $n$}\label{sec5}
We have known that the Standard Model is not the prefect and ultimate description of nature, not only because it does not describe Gravity, but also there are other problems inside it, like the problems of the existence of neutrino masses and the strong CP, which tells us that it is necessary to extend the Standard Model \cite{bib28}. But unfortunately, a successful ``grand unified model" of the four fundamental forces has not yet been established. In all the attempts to extend the Standard model, one of them is to modify the assumptions in special relativity, that is the violation of Lorentz invariance, which assumes that the Lorentz symmetry maybe violated in the ultrahigh energy scale.

Although the extensions of the Standard Model doesn't predict or require Lorentz invariance violation (LIV), but obviously and undoubtedly, the observation of LIV effects would towards to new physics beyond the Standard Model. The deformed or doubly special relativity (DSR) is exactly this class of theories that implement a modified set of principles of Lorentz invariance:
\begin{enumerate}[(i)]
    \item All the inertial frames are equivalent.

    \item Besides the constant speed of light, a maximum energy, $E_{L V}$, is also a universal constant.

    \item At low energy scale, that is in the limit $E / E_{L V} \rightarrow 0$ the speed of a photon returns to a universal constant $\boldsymbol{c}$.
\end{enumerate}

As a result of the above assumptions, the invariant of energy and momentum is modified as \cite{bib07,bib08}
\begin{equation}
    \label{eq32}
E^{2} f^{2}\left(E / E_{L V}\right)-\boldsymbol{p} \cdot \boldsymbol{p }g^{2}\left(E / E_{L V}\right)=m^{2}
\end{equation}

Where in the limit $E / E_{\mathrm{LV}} \rightarrow 0, f$ and $g$ are tend to 1 .

It should be noted that, the specific form of $f$ and $g$ has not yet been determined by strong experiments or powerful theories \cite{bib29}, and Eq. (\ref{eq1}) as a specific form of Eq. (\ref{eq32}) is not a consequence of any particular models of quantum gravity \cite{bib29}. It is just a rather simple model used to test LIV. The other notable problem in Eq. (\ref{eq32}) is that what does $E_{L V}$ stand for, and for this question there are usually three confused
answers:
\begin{enumerate}[(1)]
    \item  Planck scale $E_{P l}$, Where $E_{P l}=\sqrt{\hbar c^{5} / G}$. Planck and some physicists believe that when the strength of electromagnetic interaction and gravitational interaction are almost in the same order of magnitude, a new theory maybe emerge.

    \item $E_{q g}$, characterizing the scale of the quantum gravity theory, at which the effects of a quantum gravity theory become significant.

    \item $E_{L V}$, characterizing the scale of Lorentz violation, at which the effects of LIV become significant.
\end{enumerate}

Generally, many quantum gravity theories, which predict various LIV phenomena, are expected to work at the Planck scale $E_{P l}$, so the scale $E_{q g}$ is considered to be near the Planck scale $E_{P l}$. But their specific values of energy (along with the scale $E_{L V}$ ) need to be determined by experiments.

The above introduced the background of the deformed or double special relativity, and more details can be seen in Ref. \cite{bib29,bib30,bib31,bib32}. In this paper, the motivation for introducing parameter $n$ is the same as that for DSR, that is, seeking a possible form of Lorentz invariance violation, which may result in new physics beyond the Standard Model. And another motivation is that currently, there are no unified mathematical formal descriptions of DSR \cite{bib29}. In such a class of theories of DSR, a specific form of the modified dispersion relation regardless of the mechanisms is assumed first and then discuss the phenomena caused by it, including how the speed of light changes (LIV effects predict the local speed of light is variable). Different from the start point of DSR, the start point of this paper is first assuming a possible form of variable speed of light, namely discussing the functions of $n$, then naturally obtained the corresponding modified dispersion relation. Coincidentally, unlike the DSR model, which introduces a constant as the maximum energy of all particles in advance, this paper shows that a particle's energy can also have a limit just by taking a specific function of the variable speed of light. And it can also be seen that the obtained maximum energy of particle in this paper is nothing to do with the Planck Energy or the quantum gravity, but relates to the scale of Lorentz violation.

However, the greatest difficulty currently encountered is the lack of the results of a large number of experiments, which can guide the theory or screen the theory. Although the Lorentz symmetry has been rigorously tested at low or medium energy scales, it remains to be seen whether it holds at ultrahigh energy scale. In particular, for astrophysical measurements, even small LIV effects can be accumulated to a measurable level over cosmological distance propagation, which is the third motivation why we introduced a parameter $n$ to characterize the possible variation of the speed of light. In the later chapter, we will discuss the possible LIV effects caused by the introduced parameter $n$ from astrophysical observations. And if possible, the LIV effects can serve as a candidate explanation of some astrophysical phenomena.

\section{LIV effects}\label{sec6}
In recent years, with the rapid development of astronomical technology, we have obtained a large number of astronomical observation data, which have one outstanding advantage in the detection of LIV effects, as cosmological distance propagation can accumulate small LIV effects to a measurable level. Next we will discuss two typical astronomical phenomena, namely the gamma-ray burst events and photon annihilation events, in which some data of the observations in recent years have puzzled one and thus leading to some controversial views.

\subsection{Arrival time delay of high-energy photons}
According to Einstein's relativity, the local speed of light is a constant in vacuum. However, it is speculated that the effect of modified dispersion relation may bring a tiny correction to the speed of light of the order $E / E_{\mathrm{LV}}$, where $E$ is the photon energy. But it is very difficult to measure the light speed variation by ordinary experiments on Earth as such a variation of $c$ is extremely small. One approach to solve this problem is to focus on particles with ultrahigh energy from far away astrophysical objects. Amelino-Camelia et al. first suggested detecting light speed variation due to the Lorentz invariance violation (LIV) from gamma-ray bursts (GRBs) \cite{bib33,bib34}, as Gamma-ray bursts are extremely energetic and rather quick processes in the universe. So next we will use the GRB events to investigate the dispersion relation derived in this paper.

In many literature, such as Ref. \cite{bib15,bib16,bib35,bib36}, Eq. (\ref{eq1}) is simplified to the first order (i.e., $\chi_{2}=0$ ) or the second order (i.e., $\chi_{1}=0$ ) as the leading modification of the Lorentz violation model. Then the speed of light derived by $\boldsymbol{v}(E)=\partial E / \partial \boldsymbol{p}$ corresponding to the above three forms of modified dispersion relation can be written as
\begin{equation}
    \label{eq33}
\left\{\begin{array}{l}
\boldsymbol{v}_{1}(E)=\frac{\partial E}{\partial \boldsymbol{p}}=\frac{1}{1-\frac{Q-1}{\ln Q}+2 \frac{Q-1}{\ln Q} \frac{E}{E_{Q G}}} \boldsymbol{c} \\
\boldsymbol{v}_{2}(E)=\frac{\partial E}{\partial \boldsymbol{p}}=\frac{\sqrt{1+\chi_{1} \frac{E}{E_{L V}}}}{1+\frac{3}{2} \chi_{1} \frac{E}{E_{L V}}} \boldsymbol{c} \\
\boldsymbol{v}_{3}(E)=\frac{\partial E}{\partial \boldsymbol{p}}=\frac{\sqrt{1+\chi_{2} \frac{E^{2}}{E_{L V}^{2}}}}{1+2 \chi_{2} \frac{E^{2}}{E_{L V}^{2}}} \boldsymbol{c}
\end{array}\right.
\end{equation}

Where $\boldsymbol{v}_{1}(E)$ corresponds to Eq. $(27), \boldsymbol{v}_{2}(E)$ and $v_{3}(E)$ correspond to the first and second order approximation of Eq. (\ref{eq1}) respectively.

For low energy photon (i.e., $E / E_{\mathrm{LV}} \sim 0$ or $E / E_{\mathrm{QG}} \sim 0$ ), Eq. (\ref{eq33}) can be written as
\begin{equation}
    \label{eq34}
\left\{\begin{array}{l}
\boldsymbol{v}_{1}(E) \approx\left(1-2 \frac{Q-1}{\ln Q} \frac{E}{E_{Q G}}\right)  \boldsymbol{c} \\
\boldsymbol{v}_{2}(E) \approx\left(1-\chi_{1} \frac{E}{E_{L V}}-o\left(\frac{E^{2}}{E_{L V}^{2}}\right)\right)  \boldsymbol{c} \\
\boldsymbol{v}_{3}(E) \approx\left(1-1.5 \chi_{2} \frac{E^{2}}{E_{L V}^{2}}\right)  \boldsymbol{c}
\end{array}\right.
\end{equation}

Note that, in some literature, such as Ref. $[15,16]$, the $\chi_{1} / E_{\mathrm{LV}}$ in Eq. (\ref{eq34}) is redefined as $1 / E_{\mathrm{LV}, 1}$ and $\chi_{2} / E_{\mathrm{LV}}^{2}$ is redefined as $1 / E_{\mathrm{LV}, 2}^{2}$. It can be seen from Eq. (\ref{eq34}) that the formula of $\boldsymbol{v}_{1}(E)$ is similar to the formula of $\boldsymbol{v}_{2}(E)$, which is consistent with Eq. (\ref{eq29}) (note that $Q \sim 0$ has been used). Therefore, the form of the first order approximation of Eq. (\ref{eq1}) is approximately equivalent to Eq. (\ref{eq16}) or Eq. (\ref{eq27}) for low energy photons.

In GRB event, the observed arrival time lag between light curves in different energy bands are considered to be comprised of two contributions \cite{bib37,bib38}
\begin{equation}
    \label{eq35}
\Delta t_{o b s}=\Delta t_{L I V}+\Delta t_{\mathrm{int}}
\end{equation}

Where $\Delta t_{\text {int }}$ is the intrinsic time lag between the emission of photon of a particular energy and the lowest energy photon from the GRB, and $\Delta t_{\mathrm{LIV}}$ is the time lag due to Lorentz violation.

$\Delta t_{\text {int }}$, which depends on the internal dynamics of the GRB itself and makes a big difference to the uncertainty of $\Delta t_{\mathrm{obs}}$, was formulated by certain model. In this paper we will adopt the following formula for $\Delta t_{\text {int }}$, as it is claimed to fit the data from GRB 190114C well \cite{bib38}.
\begin{equation}
    \label{eq36}
\Delta t_{\text {int }}(E)=\tau\left[\left(\frac{\varepsilon_{0}}{1 \mathrm{keV}}\right)^{-\alpha}-\left(\frac{E}{1 \mathrm{keV}}\right)^{-\alpha}\right]
\end{equation}

where $\tau$ and $\alpha$ are the free parameters.

If we re-investigated the GRB $190114 \mathrm{C}$ event in Ref. \cite{bib38}, then $\varepsilon_{0}=12.5 \mathrm{keV}$, which is the median value of the lowest reference energy band $(10-15 \mathrm{keV})$.

Considering the cosmological time dilation, the LIV-induced time lag between high-energy and low energy photons is given as following \cite{bib15,bib16,bib38}
\begin{equation}
    \label{eq37}
\Delta t_{L I V}=\left(\left.\frac{\boldsymbol{v}_{i}(E)}{ \boldsymbol{c}}\right|_{E=E_{\text {high }}}-\left.\frac{\boldsymbol{v}_{i}(E)}{ \boldsymbol{c}}\right|_{E=E_{\text {low }}}\right) K
\end{equation}
where $i=1,2,3$, and $\boldsymbol{v}_{i}$ corresponds to Eq. (\ref{eq34}). $K$ is a factor that \cite{bib15,bib16,bib38}
\begin{equation}
    \label{eq38}
\left\{\begin{array}{l}
K_{2,3}=\frac{1}{H_{0}} \int_{0}^{z} \frac{\left(1+z^{\prime}\right)^{n_{o}}}{\sqrt{\Omega_{M}\left(1+z^{\prime}\right)^{3}+\Omega_{\Lambda}}} d z^{\prime} \\
K_{1}=K_{2}
\end{array}\right.
\end{equation}
where $K_{1}, K_{2}, K_{3}$ corresponds to $\boldsymbol{v}_{1}, \boldsymbol{v}_{2}, \boldsymbol{v}_{3}$, respectively. $H_{0}$ is the today's Hubble constant, $\Omega_{\mathrm{M}}, \Omega_{\Lambda}$ are cosmological parameters.

\begin{figure}
    \centering
    \includegraphics[width=0.8\linewidth]{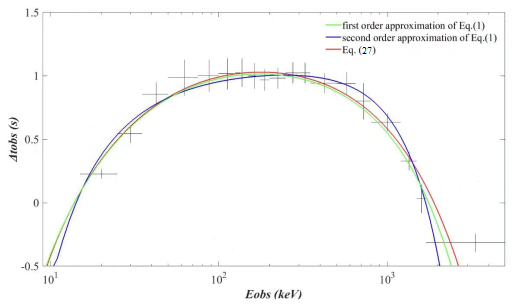}
    \caption{The observed time lag between the softest energy band $(10-15 \mathrm{keV})$ and any other high energy bands. The curves corresponding to the first order and the second order approximation of Eq. (\ref{eq1}) come from Ref. [38].}
    \label{fig02}
\end{figure}

With the above observed time lag between the reference energy band (the value corresponding to GRB $190114 \mathrm{C}$ is $10-15 \mathrm{keV}$ ) and the 19 high-energy bands in Ref. \cite{bib38}, we can simultaneously fit the free parameters $\Theta=\left(E_{\mathrm{LV}}, \tau, \alpha^{i}\right)$ (here in order to be consistency with the usual literature, we redefined that $E^{1}{ }_{\mathrm{LV}}=\ln \mathrm{Q} /[2(\mathrm{Q}-1)] E_{\mathrm{QG}}$, $E^{2}{ }_{\mathrm{LV}}=E_{\mathrm{LV}} / \chi_{1}, E_{\mathrm{LV}}^{3}=E_{\mathrm{LV}} / \mathrm{sqrt}\left(\chi_{2}\right)$, and $E_{\mathrm{LV}}^{i}(i=1,2,3)$ corresponds to $\left.\boldsymbol{v}_{i}(i=1,2,3)\right)$ involved in the above three models in Eq. (\ref{eq34}) with a Beyesian approach \cite{bib39}. The constraint results were shown in Tab. 1, and the theoretical curves of spectral lags corresponding to the median values along with the confidence bands of $2 \sigma$ error bars was displayed in Fig. \ref{fig02}.

\begin{table}[h]
    \centering
    \caption{The constraint results at $2 \sigma$ confidence level, the reduced chi-squared, and the Bayes factor for different models}\label{tab01}
    \begin{tabular}{cccc}
\toprule
Model (parameters) & $2 \sigma$ bounds \cite{bib38} & $\chi^{2} /$ dof & Bayes factor \\
\midrule
No LIV $(\tau, \alpha)$ & $353.42_{-263.87}^{+357.66} \mathrm{GeV}, 2.46_{-0.55}^{+0.28} \mathrm{~s}$ & $384.63 / 17$ & $/$ \\

$\left(E^{1}_\mathrm{LV}, \tau, \alpha\right)$ in Eq. $(27)$ & $14.49_{-0.13}^{+0.12} \mathrm{GeV}, 10.61_{-5.52}^{+22.70} \mathrm{~s}, 0.84_{-0.41}^{+0.50}$ & $7.09 / 16$ & $5.97 \times 10^{75}$ \\

$\left(E^{2}_\mathrm{LV}, \tau, \alpha\right)$ in the first order & \multirow{2}*{$14.49_{-0.13}^{+0.12} \mathrm{GeV}, 10.61_{-5.52}^{+22.70} \mathrm{~s}, 0.84_{-0.41}^{+0.50}$ }& \multirow{2}*{$8.39 / 16$}& \multirow{2}*{$1.08 \times 10^{76}$ }\\

approximation of Eq. (\ref{eq1})  & & &\\

$\left(E^{3}_\mathrm{LV}, \tau, \alpha\right)$ in the second  & \multirow{2}*{$6.00_{-0.06}^{+0.06} \mathrm{GeV}, 22.16_{-13.77}^{+45.43} \mathrm{~s}, 1.12_{-0.42}^{+0.46}$ }& \multirow{2}*{$7.36 / 16$} & \multirow{2}*{$6.21 \times 10^{75}$ }\\
order approximation of Eq. (\ref{eq1}) &  &  &  \\

\bottomrule
    \end{tabular}
\end{table}

Tab. \ref{tab01} shows that for GRB 190114C, the fitting results of Eq. (\ref{eq27}) are slightly better than that of the two models corresponding to the first or second order approximation of Eq. (\ref{eq1}).

\subsection{Threshold of Photon Annihilation Reaction}
In recent years, astronomical observatory have reported that some ultrahigh energy photon events from the universe are detected, such as the Large High Altitude Air Shower Observatory (LHAASO) reported that in GRB 221009A event (located at around $\mathrm{RA}=288.282$ and $\mathrm{Dec}=19.495$ ), more than 5000 very high energy photon events with energies larger than $500 \mathrm{GeV}$ are detected, including photons with the
energy of about $18 \mathrm{TeV}$ \cite{bib40}, and in 2021, LHAASO reported that a large number of $\gamma$-ray photons with energies above $100 \mathrm{TeV}$ from cosmic accelerators within the Milky Way are detected \cite{bib41}, even including the most high energetic $\gamma$-ray at about 1.4 PeV. However, the events of such detectable ultrahigh-energy photons have puzzled some physicists, for that if following the standard knowledge of particle physics and universe, such high energies photons would probably be undetectable. It's well known that in the universe there exists an energy threshold corresponding to the photon annihilation reaction $\gamma+\gamma_{\mathrm{b}} \rightarrow \mathrm{e}^{+}+\mathrm{e}^{-}$that can prevent cosmic photons with energy higher than the energy threshold from traveling large distances in the universe. According to the special relativity, the energy threshold corresponding to high-energy photon is \cite{bib42,bib43}
\begin{equation}
    \label{eq39}
E^{\prime} \geq E_{t h}=\frac{m_{e}^{2}}{\varepsilon_{b}}
\end{equation}

Where $E^{\prime}$ represents the energy of the gamma-ray, $m_{e}$ is the mass of electrons. $\varepsilon_{b}$ represents the energy of photon from background radiation fields, and if the cosmic microwave background (CMB) is chosen as the background radiation field, the mean energy of $\varepsilon_{b}$ is about $6.35 \times 10^{-4} \mathrm{eV}$, which resulting in the threshold energy $E_{\mathrm{th}}{ }^{\mathrm{CMB}} \approx$ $411 \mathrm{TeV}$.

There are many possible reasons to explain why the energy of detectable photons is higher than the energy threshold, for example, one possible explanation is that the photons with energy at about $1.4 \mathrm{Pev}$ come from sources that are not very far away from the earth such that they did not suffer from the photon absorption very severely \cite{bib41,bib44}. However, LIV effects is also considered to be a candidate explanation for such ultrahigh energy events \cite{bib45,bib46}. That is, based on the conclusions of Ref. \cite{bib46}, the dispersion relation in Eq. (\ref{eq16}), which deviates from the Lorentz model, can result in Eq. (\ref{eq39}) modified as
\begin{equation}
    \label{eq40}
E^{\prime} \geq E_{t h} \approx \frac{m_{e}^{2}}{\varepsilon_{b}}+\frac{1}{4 \varepsilon_{b}}\left(\frac{Q-1}{\ln Q}\right)^{2} \frac{E^{\prime^{4}}}{E_{Q G}^{2}}+\frac{1}{2 \varepsilon_{b}} \frac{Q-1}{\ln Q} \frac{E^{\prime^{3}}}{E_{Q G}}
\end{equation}

From Eq. (\ref{eq40}) we can obtain the critical photon energy $E^{\prime}=E_{\text {cr }}$ that makes the $\varepsilon_{b}$ have a global minimum, and for photons with energy above the critical point, the reaction threshold will increase with the energy of gamma-ray photon energy increases \cite{bib45}. Since the density of photons in background radiation field decreases with energy increases, there are less background photons to interact with the energetic photon with energy above the critical energy, therefore, the energetic photons have the potential to travel large distance and reach the earth.

The critical photon energy $E_{\mathrm{cr}}$ can be obtained by requiring $\partial \varepsilon_{b} / \partial E^{\prime}=0$
\begin{equation}
    \label{eq41}
E_{c r} \approx\left(m_{e}^{2} \cdot \frac{\ln Q}{Q-1} E_{Q G}\right)^{1 / 3}
\end{equation}

Note that the above we takes the first order approximation. Here if we substitute the obtained parameters in Sect. 6.1 into Eq. (\ref{eq41}) then it can be obtained that $E_{c r} \approx 0.55 \mathrm{TeV}$, and correspondingly, the background photon threshold energy reaches its minimum
$\left.\varepsilon_{b}\right|_{\text {min }} \approx 0.71 \mathrm{eV}$, which is over 3 orders of magnitude the mean energy of the current $\mathrm{CMB}$ radiation. It indicates that the $\mathrm{CMB}$ photons with energy in general lower than this minimal threshold can never interact with the energetic gamma-rays.

In summary, as stated in many literature \cite{bib42,bib43,bib44,bib45,bib46} that, the violation of Lorentz model corresponding to the subluminal case can increase the energy threshold of photon annihilation reaction, here in this paper we obtained this conclusion again. With more data on the energetic photon events in the future, we expected that it can constrain some Lorentz violation models and arrive at a more concrete conclusion.

\section{Conclusions}\label{sec7}
To this day, physicists are still trying to use ultrahigh energy events to test the Lorentz model, and an important feature for the violation of Lorentz model is that the local speed of light is variable. In this paper we present a parameter $n$ to characterize the violation of Lorentz model. And similar to the DSR, we try to construct an expression for $n$ to make the particle's energy have a limit. However, due to the lack of physical evidence or rules at present, it seems difficult to determine the specific form of $n$. But under the constraints of various experiments, and as well as a matter of preference, we prefer to choose a simpler form as the eligible function of $n$ in this paper, that is refer to a function with one undetermined parameter $Q$. Thus we can test the value of $Q$ to indirectly check the energy scale of violation of the Lorentz model (if $Q \equiv 0$ then $n \equiv 1$, correspondingly, Eq. (\ref{eq10}) returns to the case in Lorentz model).

But with regret, it seems that the existing experiments still cannot specific the value of $Q$. The key point is that it is currently difficult to increase the energy of particles to be a larger scale due to the limit of technical means. As a general rule, what we can still do is we can obtained the upper bound of $Q$. For example, based on T. Alvager's experiments [27], one can obtain that $Q<(1 / e)^{10^{6}}$. In addition, Eq. (\ref{eq29}) or Eq. (\ref{eq30}) can also help us obtain the upper bound of $Q$ indirectly in some experiments aiming to test the rainbow model.

In many literature, the researcher used the simplified formula of Eq. (\ref{eq1}), i.e., the first or second order approximation of Eq. (\ref{eq1}), to investigate the GRB events and in turn constrain the parameters involved in the model. In this paper we used Eq. (\ref{eq27}), which can be regarded as a special case of Eq. (\ref{eq1}), to investigate the photon arrival time lag in GRB 190114C event, and found that it fit the current data slightly better than the other two models. And in this paper we also obtained the conclusion that the modified dispersion relation in Eq. (\ref{eq16}) can increase the energy threshold of photon annihilation reaction, which is same as the conclusion in the other literature. But it seems not to enough. We look forward to further testing the LIV effects and screening these LIV models in the future ultrahigh energy experiments, as different LIV models predict different phenomena, which can also be seen from Eq. (\ref{eq33}) that the higher the particle energy, the greater the difference in particle's behavior predicted by the above three models.

\bibliographystyle{elsarticle-num-names}
\bibliography{cas-refs}

\end{document}